\documentstyle[12pt,aps,epsfig]{revtex}

\newcommand\simlt{\lower.5ex\hbox{$\; \buildrel < \over \sim \;$}}
\newcommand\simgt{\lower.5ex\hbox{$\; \buildrel > \over \sim \;$}}

\begin{document}

\title{Gravitational radiation from a torus around a black hole}
\author{Maurice H.P.M. van Putten, MIT, Cambridge, MA 02139-4307}

\maketitle
\mbox{}\\
\mbox{}\\
\begin{abstract}
  Long gamma-ray bursts (GRBs) from rapidly spinning black hole-torus 
  systems may represent hypernovae or black hole-neutron star coalescence. 
  We show that the torus may radiate gravitational radiation powered by the
  spin-energy of the black hole in the presence of non-axisymmetries.
  The coupling to the spin-energy of the black hole is due to 
  equivalence in poloidal topology to pulsar magnetospheres. 
  Results calculated in the suspended accretion state indicate that
  GRBs are potentially the most powerful LIGO/VIRGO burst-sources in the 
  Universe, with an expected duration of 10-15s 
  on a horizontal branch of 1-2kHz in the $\dot{f}(f)-$diagram.
\end{abstract}
\mbox{}\\

Cosmological gamma-ray bursts (GRBs) are the most enigmatic events 
in the Universe. Their emissions are characteristically non-thermal in
the range of a few hundred keV. The BATSE catalogue 
shows a bi-modal distribution of GRB durations, with short bursts of 
about 0.3s and long bursts of about 30s \cite{kou93}. 
The afterglow phenomenon -- broad band secondary emissions generally
towards lower energies -- has revolutionized our understanding of GRBs 
as internal shocks in baryon poor (leptonic) outflows and external 
shocks in interaction with the interstellar medium (see \cite{pir98}).
The notion of GRB-emissions from shocks a certain distance away from the 
source \cite{ree92} gracefully circumvents the original compactness
problem.
The energetics and rapid temporal variability observed in GRBs suggest an 
association with energetic compact sources. Notable candidates are hypernovae 
in star-forming regions 
\cite{pac97,blo00} and black hole-neutron star coalescence \cite{pac91}. 
Hypernovae represent a variant on failed supernovae \cite{woo93}, 
produced in core-collapse of rapidly rotating, strongly magnetized massive 
stars.
Recently, it has been suggested \cite{bro00} that relics of 
GRBs might be found in soft 
X-ray transients which show chemical abundances in 
$\alpha-$nuclei, such as GRO J1655-40
\cite{isr99} and V4641 Sgr \cite{oro01}. This potential GRB/SXT connection by
hypernovae \cite{bro00} could become a relevant factor in GRB-phenomenology, 
next to their cosmological origin and bi-modal distribution. 

Catastrophic events from high-angular momentum compact sources such as 
hypernovae or black hole-neutron star coalescence are expected to result in black hole 
plus disk or torus systems (see \cite{mvp01} for a review). 
Observational support for black hole plus disk/torus-systems presently consists of 
the following. Short/long bursts may be identified with hyperaccretion/suspended-accretion 
onto slowly/rapidly spinning black holes \cite{mvp01c}. 
This also points towards a positive
correlation between fluence and the spin-rate of the black hole, in agreement with
the distinct values of $<V/V_{max}>$ for long and short bursts \cite{kat96}.
This suggests that short bursts may feature afterglows
(cf. the 2s event GRB 000301C \cite{jen01}). 
Baryonic winds emanating from the torus may be powerful when derived in
suspended accretion from the spin-energy of the black hole, which may
be the input to collimation \cite{lev00,mvp01e}. 
The high-mass ejections onto a companion star in the hypernova scenario
of \cite{bro00} can derive from the spin-energy of the
central black hole by a safe margin.

In this {\em Letter}, we focus on gravitational radiation from the torus powered by the
the spin-energy of the black hole. We shall find that this represents a major fraction
of the black hole-luminosity, emitted as presently ``unseen" emissions,
whenever the torus becomes non-axisymmetric. True calorimetry of gamma-ray bursts, therefore,
may be obtained from measuring the fluence in these gravitational-wave emissions \cite{mvp01e}.
The remainder will be emitted in various ways, such as Poynting flux-winds, baryonic
collimating winds and, when sufficiently hot, neutrino emissions.
Note that this prediction for as yet unseen emissions invalidates
the often stated suggestion that the total energy budget is generally reduced for beamed
emissions, in the case of black hole-inner engines with rapid spin.
Furthermore, the emissions from the torus in low-frequency radio-waves (modulated) are of 
potential interest to the planned Low Frequency Radio Antenna (LOFAR) and the Square
Kilometre Array (SKA), suggesting
to consider correlated LIGO/VIRGO-LOFAR/SKA searches.

Gravitational radiation from a torus features several
aspects which suggest considering long GRBs as potential sources for LIGO/VIRGO.
Namely, the torus is strongly coupled to the spin-energy of the black hole; 
lumpiness in the torus will produce gravitational radiaton at twice the
Keplerian angular frequency, i.e., in the range of 1-2kHz; the emission in 
gravitational radiation should dominate over emissions in radio waves
(see young pulsars \cite{sha83}); the true rate of GRBs should be frequent
as inferred from their beaming factor $520\pm85$ \cite{frai01}. Note that gravitational 
wave-emissions from a torus in suspended accretion is powered by the spin-energy of the black 
hole and, hence, is distinct from gravitational wave-emissions from spiral-in in neutron 
star-neutron star mergers \cite{nar92,koc93} or by fragmentation in collapse towards 
supernovae \cite{bon95}.

The torus is likely to possess dynamical and, potentially, radiative instabilities. 
A geometrically thick torus is consistent \cite{mvp01e} with the recent indication that GRBs may be
standard \cite{frai01}. A thick torus is generally unstable \cite{pap84}. 
If the torus reaches an appreciable mass fraction of the central black hole, it will be
unstable to self-gravity (see \cite{woo96}). 
Similar to rapidly rotating neutron stars,
the torus may be subject to the Chandrasekhar-Friedman-Shutz instability.
Since lumps of matter radiate preferentially on inner orbits,
a quadrupolar, radial deformation of the torus might also be radiatively unstable.
It would be of interest to study these radiative instabilities in further detail. 
We note that some of the QPOs in accretion disks in X-ray binaries have been attributed
to general relativistic effects in orbital motions \cite{ste00}.

An equivalence in poloidal topology to pulsar magnetospheres shows a high
incidence of black hole-luminosity into the torus, when magnetized by
the remnant flux from the progenitor star --
a massive star in hypernovae or a neutron star in coalescence onto a black hole.
The black hole thus surrounded by a torus magnetosphere will adjust to its lowest energy
state by developing an equilibrium magnetic moment
\cite{mvp01} 
\begin{eqnarray}
\mu^e_H\approx aBJ_H,
\label{EQN_B}
\end{eqnarray}
where $B$ denotes the average poloidal magnetic field in the vicinity of the
black hole, and $a=J_H/M$ the specific angular momentum of a black hole with
angular momentum $J_H$ and mass $M$. This equilibrium magnetic moment
maintains an essentially maximal horizon flux \cite{lee01}. It serves to
preserve strong coupling to the torus magnetosphere and, through it,
to the surrounding matter. The latter follows by equivalence in poloidal
topology to pulsar magnetospheres \cite{mvp99,mvp01}, wherein the inner
and outer faces of the torus each correspond to a pulsar with an
appropriate angular velocity.
When the black hole
spins sufficiently rapidly, a state of suspended-accretion may result
(\cite{mvp01c} and below), wherein the magnetic moment (\ref{EQN_B}) can support
open magnetic field-lines to infinity. The latter may account for the
beamed outflows of baryon poor jets along the axis of rotation
\cite{mvp00,mvp01}. However, the equivalence in poloidal topology to pulsar
magnetospheres indicates that the black hole-luminosity $L_T$
onto the torus {\em far} exceeds such luminosity $L_p$ into baryon poor
jets \cite{mvp01e}.

Estimates of the various emissions from the torus can be obtained in a
suspended-accretion state \cite{mvp01c}.
Here, the emissions from the torus are replenished by spin-up Maxwell stresses on the inner face,
through the magnetic connection to the black hole. This operates by 
equivalence in poloidal topology to pulsar 
magnetospheres: the pulsar which is equivalent to the inner face has an angular velocity
$-(\Omega_H-\Omega_+)$, where $\Omega_H$ denotes the angular velocity of the
black hole and $\Omega_+$ denotes the angular velocity of the inner face.
The inner face of the torus thus receives a spin-up torque
(adapted from \cite{gol69,tho86,mvp01c})
\begin{eqnarray}
\tau_+=(\Omega_H-\Omega_+)f_H^2A^2,
\end{eqnarray}
where $f_H$ denotes the fraction of flux which reaches the
horizon of the black hole, of the
net poloidal flux $2\pi A$ supported
by the torus. In equilibrium with the radiative losses from the torus,
a suspended-accretion state will result.

The motion of the torus subject to the powerful shear between the inner and the outer faces of the torus
remains, to leading order, Keplerian.
Some deviation away from Keplerian motion is expected, as the
competing torques tend to bring the two faces in state
of super- and sub-Keplerian motion, with positive
radial pressure which promotes a radially slender shape.
The interface separating the two faces is expected to be
unstable, which favors turbulent mixing into a state of uniform 
specific energy across the torus. Mixing enhances differential 
rotation, as may be illustrated in the Newtonian limit, which
gives rise to the angular velocity $\Omega(r)\approx\Omega_K(1-(r-a)/a)^{1/2}$
as a function of radius $r$ for a torus of major radius $a$. Compression into a more slender
shape tends to reduce differential rotation. The net result  
should be that the characteristically Keplerian decrease 
of angular velocity with radius is approximately
preserved. The inner and other faces will have, respectively, angular velocities
$\Omega_\pm\approx \Omega_K\left(1\pm 3b/4a\right),$
where $b$ denotes their radial separation.
The same trend should hold in the Kerr metric. In what follows, we will neglect
such perturbations $3b/4a$ to the Keplerian velocity distribution.

Gravitational radiation from a torus surrounding a black hole tends to dominate
radio waves of the same frequency. This is generally the case for compact systems 
(of the order of their Schwarzschild radius) in the presence of gravitationally weak 
magnetic fields. Consider a torus
with ellipticity $\epsilon$, a magnetic moment $\mu_T$ and mass $m$ in rotation
about its center of mass. Its quadrupole moments in magnetic 
moment and mass are, respectively, $\epsilon \mu$ and $\epsilon m$, which produce
luminosities (adapted from \cite{sha83}):
${L}_{EM}\approx\pi^{-1}(\Omega_T M)^4(\mu_T/M^2)^2\epsilon^2$
and 
${L}_{GW}\approx ({32}/{5})(\Omega_T M)^{10/3} (m/M)^2\epsilon^2$
in geometrical units.
These emissions may be compared with, respectively, the luminosity in radio emission
$\sim \Omega^4_p\mu_p^2/\pi$ from an orthogonal pulsar and
in gravitational-wave emissions 
$\sim ({32}/{5})(\Omega_{orb} {\cal M})^{10/3}$ 
from neutron star-neutron star binaries with 
angular velocity $\Omega_{orb}$ and chirp mass
${\cal M}=(M_1M_2)^{3/5} /(M_1+M_2)^{1/5}$ (for circular orbits).
The ratio of radio-to-gravitational wave emissions can be evaluated as
\begin{eqnarray}
{{L}_{EM}}:
{{L}_{GW}}
\sim (\Omega M)^{2/3} (E_B/M) (M/m)^2 < 1,
\label{EQN_EST}
\end{eqnarray}
e.g., when $E_B/M\sim 10^{-6}$ for the relative
energy in the magnetic field and $M/m\le 10^2$.

The suspended accretion state is described 
by equilibrium conditions for torque and energy: 
\begin{eqnarray}
\left\{
\begin{array}{rl}
\tau_+         &= \tau_-       +\tau_{rad},\\
\Omega_+\tau_+ &=\Omega_-\tau_-+\Omega\tau_{rad}+P_{d},
\end{array}
\right.
\label{EQN_B1}
\end{eqnarray}
where $P_{d}$ denotes dissipation,
$\Omega\approx\Omega_K$ a mean orbital angular frequency and
$\tau_-=A^2f_w^2\Omega_-$ denotes the torque on the outer face of the torus.
In (\ref{EQN_B1}), we neglect surface stresses due to radiation
derived from $P_d$, notably so in thermal and neutrino emissions.
The net magnetic flux $2\pi A$ supported by the torus will
partially connect to the black hole and to infinity (by Poynting-flux winds), 
respectively, with fractions $f_H$ and $f_w$.
Thus, $A\approx ab <B_\theta>$
in terms of the average poloidal component $B_\theta$ in the torus.
Generally, $f_H+f_w={1}/{2}~-~1$ with
$f_H\propto (M/a)^2$ for a radially slender torus (which may be
thick in the poloidal direction) of major radius $a$. A
remainder $1-f_H-f_w$ is inactive in closed field-lines, whose endpoints 
are both on either face of the torus. These field-lines extend to
the inner light surface and the outer light cylinder, and form 
toroidal ``bags." 
Note that for small differential rotation, we have
$(\Omega_K\tau_{rad}+P_{d})/{\Omega_K\tau_{rad}}
\approx({\Omega_+\tau_+-\Omega_-\tau_-})/\Omega_K({\tau_+-\tau_-})
\approx 2,$
in which limit the efficiency of the radiation is $50\%$.

The equilibrium conditions (\ref{EQN_B1}) are closed, by
specifying the internal stresses in the torus.
We shall assume that the two faces are coupled by magnetohydrodynamical
stresses due to radial components $B_r$ of the magnetic field.
These stresses are dissipative, by Ohmic heating and
magnetic reconnection, which will heat the torus and brings
about thermal and, possibly, neutrino emissions - with no surface stresses.
By dimensional analysis
\begin{eqnarray}
P_{d}\approx A^2_r(\Omega_+-\Omega_-)^2,~~~
A_r=ah<B_r^2>^{1/2},
\end{eqnarray}
where the second equation denotes the root-mean-square of 
the radial flux averaged over
the interface between the two faces with contact area $2\pi a h$.

The magnetic stresses on and inside the torus depend differently on the
magnetic field. While internal angular momentum transport between the two faces
is mediated by $<B_r^2>^{1/2}$,
the angular momentum transport from the black hole to
the torus is by the average $<B_\theta>$. The first
comprises the spectral density average over all azimuthal
quantum numbers $m$, whereas the second only involves $m=0$.
Indeed, the net flux through the black hole is generated
by the coroting horizon charge $q\approx <B_n>J$ in magnetostatic
equilibrium (\ref{EQN_B}) with the mean external poloidal magnetic field.
This averaging process is due to the no-hair theorem. While 
the exact ratio depends on the details of the magnetohydrodynamical 
turbulence in the torus, a conservative estimate is that $A_r/A$ 
is about the square root of the number of azimuthal modes in 
the approximately uniform infrared spectrum, which
should reach up to the first geometrical break at $m=a/b$, i.e.: 
$A_r/A\approx (a/b)^{1/2}$.
Substitution of the first into the second equation of 
(\ref{EQN_B1}) gives a 
luminosity
\begin{eqnarray}
L_{GW}\simeq\Omega_T\tau_{rad}\approx\Omega^2A^2\left[3(A_r/A)^2(b/a)-2f_w^2\right]\sim
\Omega_T^2A^2,
\label{EQN_LR}
\end{eqnarray}
where $\Omega_T=(\Omega_++\Omega_-)/2$ and $\Omega_+-\Omega_-\approx (3/2)(b/a)\Omega$,
and using $f_w<1$. Substitution of the right hand-side of (\ref{EQN_LR}) in the first
equation of (\ref{EQN_B1}) gives
\begin{eqnarray}
\frac{\Omega_T}{\Omega_H}\approx \frac{f_H^2}{1+f_H^2+f_w^2}.
\end{eqnarray}
The luminosity in gravitational waves, therefore, satisfies
\begin{eqnarray}
L_{GW}\simeq L_H/2;
\label{EQN_LG}
\end{eqnarray}
the luminosity in Poynting flux-winds is smaller by a factor $f_w^2$.
In the above, the suspended-accretion state
is facilitated by magnetohydrodynamical stresses; the results do not depend on
the details of the instabilities which give rise to
the required non-axisymmetries in the torus. It would be of
interest to study the type of instability by numerical simulations.
 
On the secular time-scale of spin-down of the black hole, the Keplerian
frequency of the torus will evolve in time.
This indicates a {\em horizontal branch} of the frequency dynamics in
the $\dot{f}(f)-$diagram \cite{mvp00d}. Lumpiness in the torus will radiate at twice the Keplerian
frequency of the torus, and hence produces 
gravitational wave-frequency of 
\begin{eqnarray}
f_{gw}(t)\sim1-2kHz/(1+z), ~~~\mbox{d}f_{gw}(t)/\mbox{d}t=\mbox{const.}
\label{EQN_A1}
\end{eqnarray} 
for canonical GRB values for a black hole-torus 
system at redshift $z$. Here, the sign of the constant follows the
change in Keplerian frequency of the torus; it depends sensitively
on the details of the magnetic black hole-to-torus coupling provided by
the inner torus magnetosphere. In particular, it depends on the detailed radial 
dependence of the horizon flux as a function of the 
major radius of the torus, which is beyond the scope of the present
analysis.
If the torus shows violent behavior,
the gravitational waves may be
episodical, and be correlated with
sub-bursts in long GRBs through modulations of the equilibrium magnetic moment (\ref{EQN_B}).
In this event,
the linear chirp in (\ref{EQN_A1}) is more likely to be indicative
of an ensemble average over bursts, rather than to hold for
individual bursts. The duration should be the intrinsic duraton of the gamma-ray burst
event, i.e., about 10-15s as inferred from the mean value of
30s in the BATSE catalogue, corrected for redshift.

The black hole-luminosity is partly directed to the torus, and partly in baryon poor outflows as input
to the observed GRBs. Most of the luminosity is into the torus, since only a small fraction
is into the jet through an open flux-tube along the axis of rotation. The open flux-tube
is described by an opening angle $\theta_H$
on the horizon and an opening angle $\theta_j$ on the celestial sphere. Generally,
$\theta_j<\theta_H$ by collimation, e.g., by baryonic collimating winds \cite{lev00}.
A model dependent estimate gives an
estimate for the mean geometrical beaming $f_b=\theta_j^2/2\simeq1/520$ (for bi-polar outflows)
and an average GRB fluence of $5\times 10^{50}$ \cite{frai01}. 
The horizon opening angle $\theta_H$ should be sufficient to account for the GRB fluence, yet 
may leave a dominant fraction of the
black hole-luminosity incident on the torus.  For a bi-polar output, we have
\cite{mvp01e}:
\begin{eqnarray}
L_{p}:L_T\simeq f_o^2,
\label{EQN_L}
\end{eqnarray}
where $f_o=\theta_H^2/4$ denotes the beaming factor of the flux-cone on the horizon.
Identifying a long GRB with the spin-down
of a rapidly spinning black hole, we have $L_p:L_T\simeq E_j:E_T$,
where $E_j=E_{GRB}/\epsilon$ denotes the energy in the jet inferred from
the observed GRB-fluence at a
(model dependent) efficiency
$\epsilon$, and $E_T\simeq \Omega_T/\Omega_H E_{rot}$ denotes the
energy input to the torus derived from the rotational energy of the black hole.
This gives an estimate $\theta_H\simeq35^o$ \cite{mvp01e}, 
which may be standard if the torus is geometrically thick \cite{mvp01e}.
Thus, the ratio (\ref{EQN_L}) is small. 

The stability of the gravitational wave-frequency is somewhat uncertain,
as it may be variable by the magnetohydrodynamical
turbulence in the torus. Nonetheless, it is of interest to consider
the possibility of encountering
a well-defined secular frequency sweep 
(upwards or downwards) in case the frequency behavior is quasi-periodic. 
In this event, a Fourier analysis suffices.
The effective amplitude then correlates with the fluence in gravitational
waves - derived from an enhancement in gain
by a factor $\sqrt{n}$, where $n$ is the number
of cycles in the emission. The effective amplitude
of the gravitational radiation of a cosmologically nearby source distance $D$ satisfies
\begin{eqnarray}
h_{eff}^{grb}\sim
\left(\frac{M}{D}\right) 
\left(\frac{E_{GW}}{M}\right)^{1/2} 
\end{eqnarray}
for a net fluence $E_{GW}$ in gravitational waves.
By (\ref{EQN_LG}), $E_{GW}$ is about one-half the fraction
$\Omega_T/\Omega_H$ of the spin-energy of the black hole, i.e.,
$E_{GW}\simeq 0.1M_\odot (M/10M_\odot)$.
A geometrical beaming factor of about $500$ \cite{frai01} gives rise to 
multiple events per year within a distance $D\sim 100$Mpc with
$h_{eff}\sim 10^{-21}$.
Combined, this points towards GRBs as potential sources for LIGO/VIRGO.

In summary, black hole-torus systems
representing hypernovae or black hole-neutron
star coalescence are candidates of LIGO/VIRGO sources of gravitational radiation.
The calorimetry of their non-thermal emissions is dominated by gravitational 
radiation from the torus, which derives from the spin-energy of the black hole.
The gravitational wave-frequency is expected to be
1-2kHz on a horizontal branch in the
$\dot{f}(f)-$diagram,
by secular evolution on the time-scale of a long burst, for a duration
of about 10-15s.
If observed, these aspects discriminate these sources from,
e.g., binary coalescence or prompt emission during a core-collapse.
Current estimates of the geometrical beaming factor indicate
that the true rate of GRBs may reach a few events per
year within a distance of 100Mpc. 

Optimal strategies for detecting these gravational wave-sources
appear to be by Fourier analysis, should be the signal be quasi-periodic.
Alternatively, LIGO/VIRGO searches might be combined with future radio
searches, such as LOFAR/SKA.

{\bf Acknowledgement.} This research is supported
by NASA Grant 5-7012, an MIT C.E. Reed Fund and a NATO
Collaborative Linkage Grant. The
author thanks S. Kulkarni and R. Weiss for 
stimulating discussions.
\newpage
\centerline{\bf Figure captions}
\mbox{}\\
\mbox{}\\
{\bf Figure 1.} Cartoon of a rapidly rotating black hole-torus system in suspended accretion.
 The black hole assumes an equilibrium magnetic moment in its lowest energy state.
 The torus magnetosphere is supported by a surrounding torus. Equivalence in poloidal 
 topology to pulsar magnetospheres indicates a high incidence of the black hole-luminosity
 on the inner face of the torus. The torus reradiates this input in gravitational radiation,
 Poynting flux-winds and, possibly, neutrino emissions. 
 [Reprinted from van Putten,
 M.H.P.M., {\em Physics Reports}, 345, 1-59 \copyright 2001,
 Elsevier B.V.]

\newpage
\mbox{}\\
\vskip1in
\mbox{}\\
\centerline{
\epsfig{file=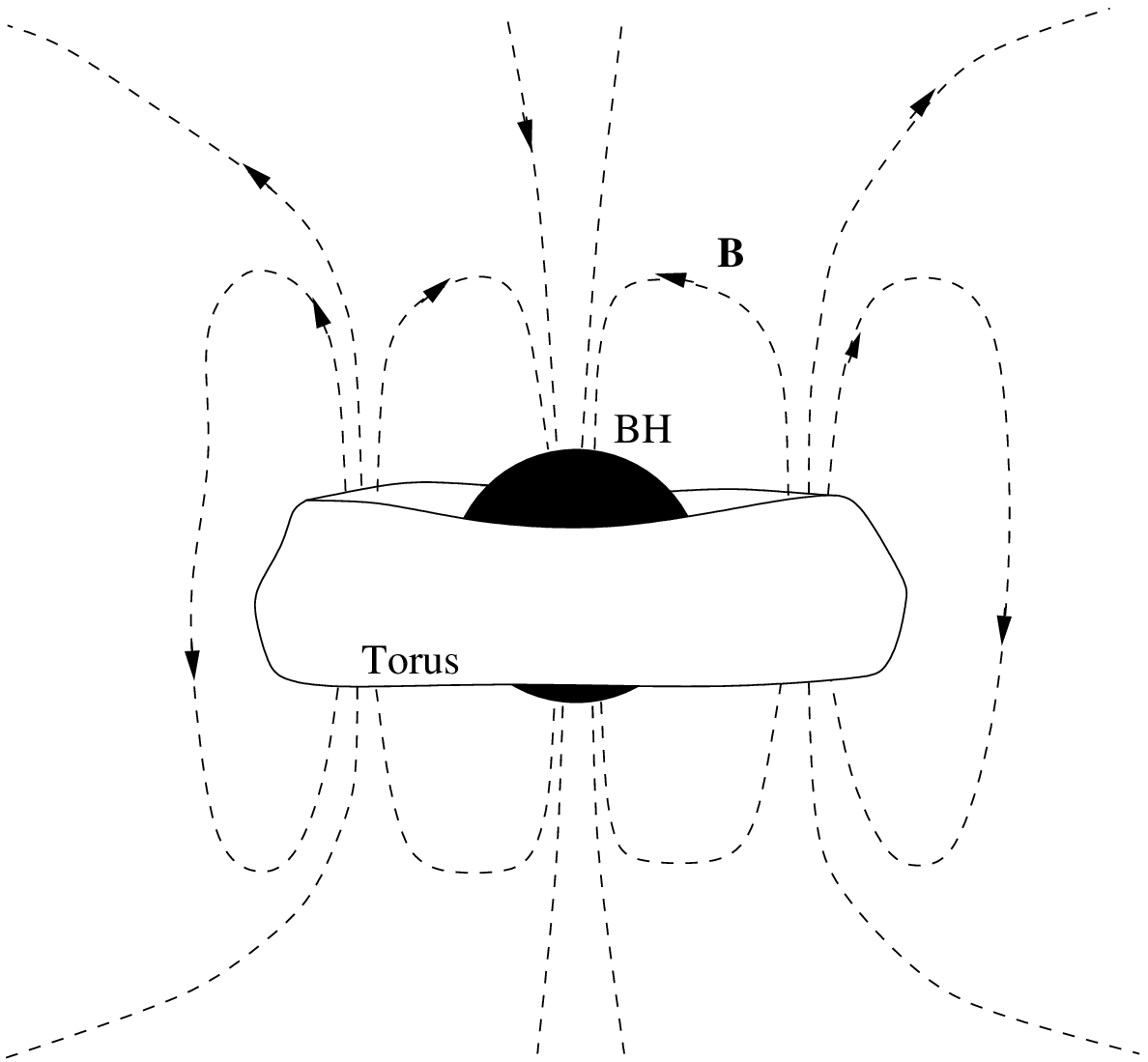}
}
\mbox{}\\
\vskip1in
{\sc FIGURE 1}
\end{document}